\documentclass[conference]{IEEEtran}
\IEEEoverridecommandlockouts

\usepackage{pgfplots}
\usepackage{tikz}
\usetikzlibrary{arrows.meta}
\pgfplotsset{compat=1.18}
\usepackage[nocompress]{cite}
\usepackage{amsmath,amssymb,amsfonts}
\usepackage{algorithm}
\usepackage{graphicx}
\usepackage{textcomp}
\usepackage{xcolor}
\usepackage{float}
\usepackage{array}
\usepackage{multirow}
\usepackage{url}
\usepackage{pgfplotstable}
\usepackage{algpseudocode}

\tikzset{
    broadcast/.style={->, thick, blue},
    aggregate/.style={<-, thick, red}
}

\def\BibTeX{{\rm B\kern-.05em{\sc i\kern-.025em b}\kern-.08em
    T\kern-.1667em\lower.7ex\hbox{E}\kern-.125emX}}

\definecolor{ieeeblue}{RGB}{0,79,149}
\definecolor{ieeeorange}{RGB}{242,147,0}
\definecolor{ieeegray}{RGB}{118,118,118}
\definecolor{ieeegreen}{RGB}{0,152,76}
\definecolor{ieeered}{RGB}{204,0,51}
\definecolor{ieeepurple}{RGB}{102,45,145}
\definecolor{ieeecyan}{RGB}{0,159,218}
\usetikzlibrary{arrows.meta,calc}
\begin{document}

\title{Semantic Communication via Federated Learning with Knowledge-Enhanced Networks and Privacy \vspace{-0.5em} \author{Hassan Touati and Rodrigo C. de Lamare 
\thanks{The authors are with the Centre for Telecommunication Studies, Department of Electrical Engineering, Pontifical Catholic University of Rio de Janeiro, Emails: touati.hassan@hotmail.com, delamare@puc-rio.br. Prof. de Lamare is also with the University of York, UK. CNPQ, CPQD, FAPERJ, and FAPESP supported this work.}}}
\maketitle

\begin{abstract}
This paper presents a novel privacy-preserving semantic communication framework that combines Federated Learning (FL) with Knowledge-Enhanced Networks (KE-NET) for distributed wireless systems. Unlike centralized semantic communication approaches that require data aggregation, our method enables multiple clients to collaboratively train a knowledge-enhanced model while keeping their data and knowledge graphs private. We develop a multi-objective federated optimization strategy that balances semantic fidelity, knowledge integration, channel robustness, and reconstruction accuracy.  Theoretical analysis reveals that federated aggregation improves effective SNR for $N$ clients through noise averaging.  Numerical results with AWGN and Rayleigh fading channels demonstrate that multi-client federated learning consistently outperforms both single-client and centralized baselines across all SNR conditions, with particularly significant gains in low-SNR scenarios.
\end{abstract}


\vspace{-1em}
\section{Introduction}

In the era of distributed computing and increasing privacy concerns, the need for secure and efficient communication systems has become paramount. Traditional communication systems, designed according to Shannon's information theory \cite{shannon1948mathematical}, focus on accurate bit-level transmission without considering the semantic meaning of transmitted information. This approach often leads to inefficient resource utilization, particularly when redundant data is transmitted. Recent advances in deep learning and natural language processing (NLP) \cite{NLP1,NLP2} have significantly improved language understanding and semantic representation capabilities, enabling new paradigms in communication that prioritize semantic meaning over bit accuracy. Semantic communication \cite{SC1,SC2} represents a fundamental shift from conventional bit-centric transmission to meaning-centric transmission.  Building upon the Transformer architecture, pioneering works like DeepSC \cite{DeepSC} have demonstrated that deep learning-enabled systems can maximize system capacity while minimizing semantic errors by recovering sentence meaning rather than individual bits, showing particular robustness in low signal-to-noise ratio environments.  More recently, GenSC \cite{GenSC} extends this paradigm by incorporating bidirectional token-level correlations through BART-like models, enabling the system to correct or fill in semantically similar tokens when corruption occurs during transmission.
Recent knowledge-enhanced approaches in semantic communication \cite{2023Know, wang2022knowledge, chen2021knowledge} have explored knowledge graph integration but with a limited scope. These methods typically employ knowledge graphs only at the receiver side as additional feature extractors, constraining their impact on overall semantic understanding.  Furthermore, existing approaches lack comprehensive multi-objective optimization frameworks that balance knowledge integration with channel robustness and semantic fidelity.
Unlike existing semantic communication methods that optimize one or two objectives independently, the proposed framework jointly optimizes semantic fidelity, knowledge consistency, channel robustness, and reconstruction accuracy within a unified federated learning framework.
In this work, we present a novel approach that combines semantic communication with Federated Learning \cite{FL1,FL2,FL3,qafl} (FL) and Knowledge Enhanced networks (KE-NET) \cite{ke-net}, addressing both data privacy challenges and effective communication across distributed systems. Unlike existing centralized semantic communication methods that require data aggregation at a central server, our approach enables decentralized model training across multiple clients who collaboratively learn a shared semantic communication model. Critically, raw data remains local throughout the training process, ensuring complete privacy preservation. The main novelty of the proposed framework lies in integrating knowledge graph enhancement at both the transmitter and receiver within a federated learning architecture, enabling semantic-aware communication across distributed clients without compromising data privacy. In particular, we develop a multi-objective federated optimization strategy that simultaneously balances semantic fidelity, knowledge integration, channel robustness, and reconstruction accuracy through gradient-based recursions with explicit expressions for each loss component. A theoretical analysis demonstrates that federated aggregation achieves inherent SNR enhancement through noise averaging, with experimental validation confirming significant performance improvements in challenging low-SNR scenarios. Simulations show the performance of the proposed approach against existing techniques. 

This paper is organized into the following sections. \ref{section2} presents the KE-NET architecture and mathematical formulations. \ref{section3} details the federated learning optimization framework with gradient expressions. \ref{section4} provides SNR enhancement analysis through federated aggregation theory. \ref{section5} describes the experimental setup and datasets, and presents experimental results and performance comparisons. \ref{section6} concludes the paper.

\vspace{-.5em}

\section{KE-NET Architecture and System Model}
\label{section2}
This section introduces the KE-NET architecture and its core components, detailing how they interact within the semantic communication framework.
\vspace{-.5em}
\subsection{KE-NET Definition and Components}
\begin{figure}[t]
\centering
\resizebox{0.48\textwidth}{!}{%
\begin{tikzpicture}[font=\scriptsize, >=Stealth, node distance=0.8cm]
  \usetikzlibrary{shadows, positioning, shapes.geometric, fit, arrows.meta}
  \tikzset{
    block/.style={draw, rounded corners=3pt, drop shadow={opacity=0.3},
      minimum width=1.2cm, minimum height=1.5cm, font=\footnotesize, align=center, 
      thick, line width=0.8pt},
    enc/.style={block, fill=blue!12, draw=blue!80!black},
    kg/.style={block, fill=green!12, draw=green!80!black},
    chan/.style={block, fill=red!12, draw=red!80!black},
    dec/.style={block, fill=orange!12, draw=orange!85!black},
    arr/.style={->, line width=1.2pt, >=Stealth, color=black},
    kgarr/.style={->, dashed, thick, line width=1.0pt, color=green!75!black},
    label/.style={font=\small, color=black!80}
  }
  
  \node[enc] (enc) at (0,0) {\textbf{Semantic}\\\textbf{Encoder}\\\textcolor{gray!70}{\scriptsize $f_{\text{enc}}(\cdot)$}};  
  \node[kg] (kg_tx) [right=0.6cm of enc] {\textbf{Knowledge}\\\textbf{Integration}\\\textcolor{gray!70}{\scriptsize $\text{GNN}(\cdot)$}};
  \node[chan] (chan) [right=0.7cm of kg_tx] {\textbf{Wireless}\\\textbf{Channel}\\\textcolor{gray!70}{\scriptsize $h, n$}};
  \node[dec] (dec) [right=0.7cm of chan] {\textbf{Semantic }\\\textbf{Decoder}\\\textcolor{gray!70}{\scriptsize $f_{\text{dec}}(\cdot, \mathcal{G})$}};
  
  \draw[arr] ([xshift=-0.5cm]enc.west) -- (enc.west);
  \node[font=\normalsize, above=0.05cm] at ([xshift=-0.6cm]enc.west) {$\mathbf{s}$};
  
  \draw[arr] (dec.east) -- ([xshift=0.5cm]dec.east);
  \node[font=\normalsize, above=0.05cm] at ([xshift=0.6cm]dec.east) {$\hat{\mathbf{s}}$};
  
  \draw[arr] (enc) -- (kg_tx);
  \draw[arr] (kg_tx) -- (chan);
  \draw[arr] (chan) -- (dec);
  
  \node[font=\normalsize, above=0.05cm, xshift=1.1cm] at ($(enc)!0.5!(kg_tx)$) {$\mathbf{h}_{\text{sem}}$};
  \node[font=\normalsize, above=0.05cm, xshift=3.4cm] at ($(kg_tx)!0.5!(chan)$) {$\mathbf{h}_{\text{ke}}$};
  \node[font=\normalsize, above=0.05cm, xshift=5.3cm] at ($(chan)!0.5!(dec)$) {$\mathbf{y}$};
  
  \node[draw, cylinder, fill=green!20, draw=green!80!black, 
        minimum width=.4cm, minimum height=0.9cm, 
        above=2.2cm of chan, thick, line width=0.8pt, drop shadow={opacity=0.3}] (kgicon) {};
  \node[font=\footnotesize, above=0.1cm of kgicon, color=green!80!black] {\textbf{Shared Knowledge Graph}};
  \node[font=\scriptsize, below=0.3cm of kgicon, color=green!80!black] {$\mathcal{G} = (\mathcal{V}, \mathcal{E})$};
 
  \node[font=\small, above=0.06cm of chan, color=red!80!black] {\textbf{AWGN/Rayleigh Fading}};
  
  \draw[kgarr] (kgicon.south west) -- (kg_tx.north);
  \draw[kgarr] (kgicon.south east) -- (dec.north);
  
  \node[font=\small, below=1cm] at ($(enc)!0.5!(kg_tx)$) {\textbf{TRANSMITTER}};
  \node[font=\small, below=0.2cm of dec, color=orange!80!black] {\textbf{RECEIVER}};
  
  \node[draw, dashed, rounded corners=5pt, 
        fit={(enc) (kg_tx) ([yshift=-0.8cm]enc.south) ([yshift=-0.8cm]kg_tx.south)}, 
        inner sep=1pt, color=blue!60, line width=0.6pt] {};
  \node[draw, dashed, rounded corners=5pt, 
        fit={(dec) ([yshift=-0.8cm]dec.south)}, 
        inner sep=1pt, color=orange!60, line width=0.6pt] {};
\end{tikzpicture}}

\caption{KE-Net architecture showing bidirectional knowledge graph integration at both transmitter and receiver.}
\label{fig:architecture}
\end{figure}
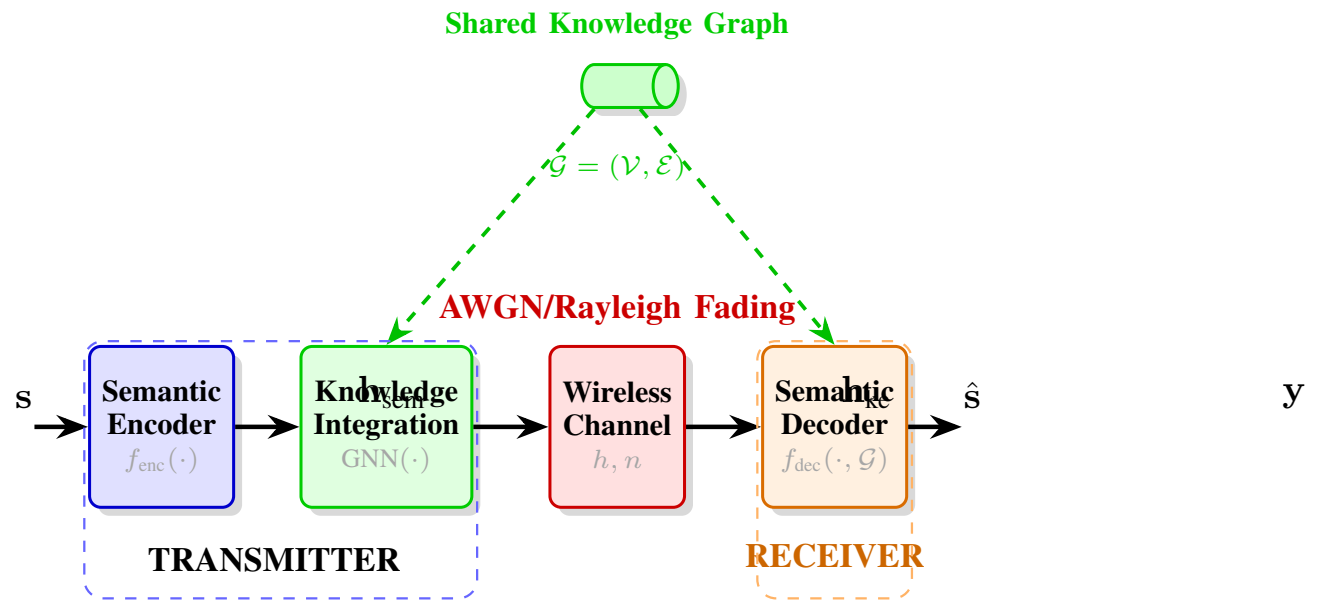

KE-NET redefines semantic communication by integrating external knowledge graphs into the encoding and decoding processes. Unlike traditional semantic communication systems that rely solely on learned representations, KE-NET leverages structured knowledge to enhance semantic understanding and improve transmission robustness under adverse channel conditions. Figure \ref{fig:architecture} illustrates the overall system architecture.
The KE-NET architecture consists of four main components:
\begin{enumerate}
    \item \textbf{Semantic Encoder}: A Transformer-based encoder that transforms input sentences into meaningful vector representations, capturing syntactic and semantic relationships.
    \item \textbf{Knowledge Integration Module}: Enhances semantic representations by incorporating relevant knowledge from external knowledge graphs (ConceptNet and WordNet), creating knowledge-aware embeddings.
    \item \textbf{Channel-Aware Transmission}: Adapts the encoded representations for robust transmission over noisy wireless channels, including power normalization and channel-specific optimization.
    \item \textbf{Semantic Decoder}: Reconstructs the original sentence meaning from the received knowledge-enhanced representations, incorporating error correction mechanisms. 
\end{enumerate}

\subsection{Mathematical Formulation}
Given an input sentence $\mathbf{s} = \{w_1, w_2, \ldots, w_L\}$, the KE-NET system operates as follows:

\textbf{Semantic Encoding:}
\vspace{-1em}
\begin{equation}
\mathbf{h}_{\text{sem}} = f_{\text{enc}}(\mathbf{s}; \theta_{\text{enc}})
\end{equation}
\textbf{Knowledge Enhancement:}

\vspace{-1em}
\begin{equation}
\mathbf{h}_{\text{ke}} = f_{\text{kg}}(\mathbf{h}_{\text{sem}}, \mathcal{G}; \theta_{\text{kg}})
\end{equation}
where $\mathcal{G} = (\mathcal{V}, \mathcal{E})$ represents the knowledge graph with vertices $\mathcal{V}$ and edges $\mathcal{E}$.

\textbf{Channel Transmission:}
\begin{equation}
\mathbf{x} = f_{\text{ch\_enc}}(\mathbf{h}_{\text{ke}}; \theta_{\text{ch}})
\end{equation}
The received signal after channel transmission is:

\vspace{-1em}
\begin{equation}
\mathbf{y} = \mathbf{h} \odot \mathbf{x} + \mathbf{n}
\end{equation}
where $\mathbf{h}$ is the channel response and $\mathbf{n}$ is additive noise.
\textbf{Semantic Decoding:}
\begin{equation}
\hat{\mathbf{s}} = f_{\text{dec}}(f_{\text{ch\_dec}}(\mathbf{y}), \mathcal{G}; \theta_{\text{dec}})
\end{equation}
\subsection{Knowledge Graph Integration}
The knowledge context $\mathbf{k} \in \mathbb{R}^{16}$ from ConceptNet and WordNet is aligned to the semantic encoder dimension ($d=512$) via projection matrices. With 8 attention heads, each head operates on dimension $d_a = 64$:
\begin{align}
    \mathbf{Q} &= \mathbf{s}_{\text{b}}\,\mathbf{W}_Q, \quad 
    \mathbf{W}_Q \in \mathbb{R}^{512 \times 64}, \\
    \mathbf{K} &= \mathbf{k}\,\mathbf{W}_K, \quad 
    \mathbf{V} = \mathbf{k}\,\mathbf{W}_V, \quad 
    \mathbf{W}_K, \mathbf{W}_V \in \mathbb{R}^{16 \times 64}.
\end{align}
Cross-attention produces the knowledge-enhanced representation:
\begin{equation}
    \mathbf{s}_{\text{k}} = \text{softmax}\left(\frac{\mathbf{Q}\,\mathbf{K}^\top}{\sqrt{64}}\right)\mathbf{V} + \mathbf{s}_{\text{b}}
    \in \mathbb{R}^{L \times 64},
    \label{eq:ke_attention}
\end{equation}
with train parameters $\theta_{\text{kg}} = \{\mathbf{W}_Q, \mathbf{W}_K, 
\mathbf{W}_V, \{\mathbf{W}_{\text{self}}^{(\ell)}, \mathbf{W}_r^{(\ell)}\}_{\ell=1}^{3}\}$
The residual connection preserves semantic content while injecting structured 
knowledge, enabling robust recovery under challenging channel conditions.

\begin{algorithm}[t]
\caption{KE-NET Training Algorithm}
\label{alg:kenet_training}
\begin{algorithmic}[1]
\Require Input dataset $\mathcal{D}$, knowledge graph $\mathcal{G}$, hyperparameters $\{\alpha, \beta, \gamma, \delta, \eta\} $ =$ \{0.4, 0.2, 0.2, 0.2, 0.0001 \}$
\Ensure Trained KE-NET model parameters $\theta^*$
\State Initialize model parameters $\theta_0$
\State Initialize optimizer parameters
\For{$\text{epoch} = 1, 2, \ldots, N_{\text{epochs}}$}
    \For{each batch $\mathcal{B} \in \mathcal{D}$}
        \State $\mathbf{h}_{\text{sem}} \gets f_{\text{enc}}(\mathbf{s}; \theta_{\text{enc}})$ \Comment{Semantic encoding}
        \State $\mathcal{G}_{\text{rel}} \gets \text{EntityLink}(\mathbf{s}, \mathcal{G})$ \Comment{Extract relevant knowledge}
        \State $\mathbf{k} \gets f_{\text{kg}}(\mathcal{G}_{\text{rel}}; \theta_{\text{kg}})$ \Comment{Knowledge embeddings}
        \State $\mathbf{h}_{\text{ke}} \gets \text{Attention}(\mathbf{h}_{\text{sem}}, \mathbf{k})$ \Comment{Knowledge integration}
        \State $\mathbf{x} \gets f_{\text{ch\_enc}}(\mathbf{h}_{\text{ke}}; \theta_{\text{ch}})$ \Comment{Channel encoding}
        \State $\mathbf{y} = \mathbf{h} \odot \mathbf{x} + \mathbf{n}$ \Comment{Channel transmission}
        \State $\hat{\mathbf{s}} \gets f_{\text{dec}}(f_{\text{ch\_dec}}(\mathbf{y}); \theta_{\text{dec}})$ \Comment{Semantic decoding}
        \State Compute loss components:
        \State \quad $\mathcal{L}_{\text{semantic}} \gets 1-\text{BLEU}(\hat{\mathbf{s}}, \mathbf{s})$
        \State \quad $\mathcal{L}_{\text{knowledge}} \gets \text{KG\_Loss}(\mathbf{h}_{\text{ke}}, \mathcal{G}_{\text{rel}})$
        \State \quad $\mathcal{L}_{\text{reconst}} \gets \|\hat{\mathbf{s}} - \mathbf{s}\|_2^2$
        \State \quad $\mathcal{L}_{\text{channel}} \gets \mathbb{E}_{\mathbf{h}}[\|\mathbf{h} \odot \mathbf{x} - \mathbf{y}\|_2^2]$
        \State $\mathcal{L}_{\text{total}} \gets \alpha \mathcal{L}_{\text{semantic}} + \beta \mathcal{L}_{\text{knowledge}} + \gamma \mathcal{L}_{\text{reconst}} + \delta \mathcal{L}_{\text{channel}}$
    \EndFor
    \If{convergence criterion met}
        \State \textbf{break}
    \EndIf
\EndFor
\State \Return $\theta^*$
\end{algorithmic}
\end{algorithm}

\section{Federated Learning Framework} 
\label{section3}

This section presents the federated learning framework for semantic communication, comparing performance across single-client and multi-client configurations under varying SNR conditions. We analyze how collaborative training through aggregation enhances model performance while preserving data privacy through decentralized learning. We remark that similar strategies could be investigated with a peer-to-peer architecture using distributed processing \cite{dce,rdrls,dmest,lrcc,als,damdc,dqa,dqarls,rdiff}.

\subsection{Federated Learning Framework Overview}

\begin{figure}[h!]
    \centering
    \includegraphics[width=0.9\linewidth, height=0.25\textheight]{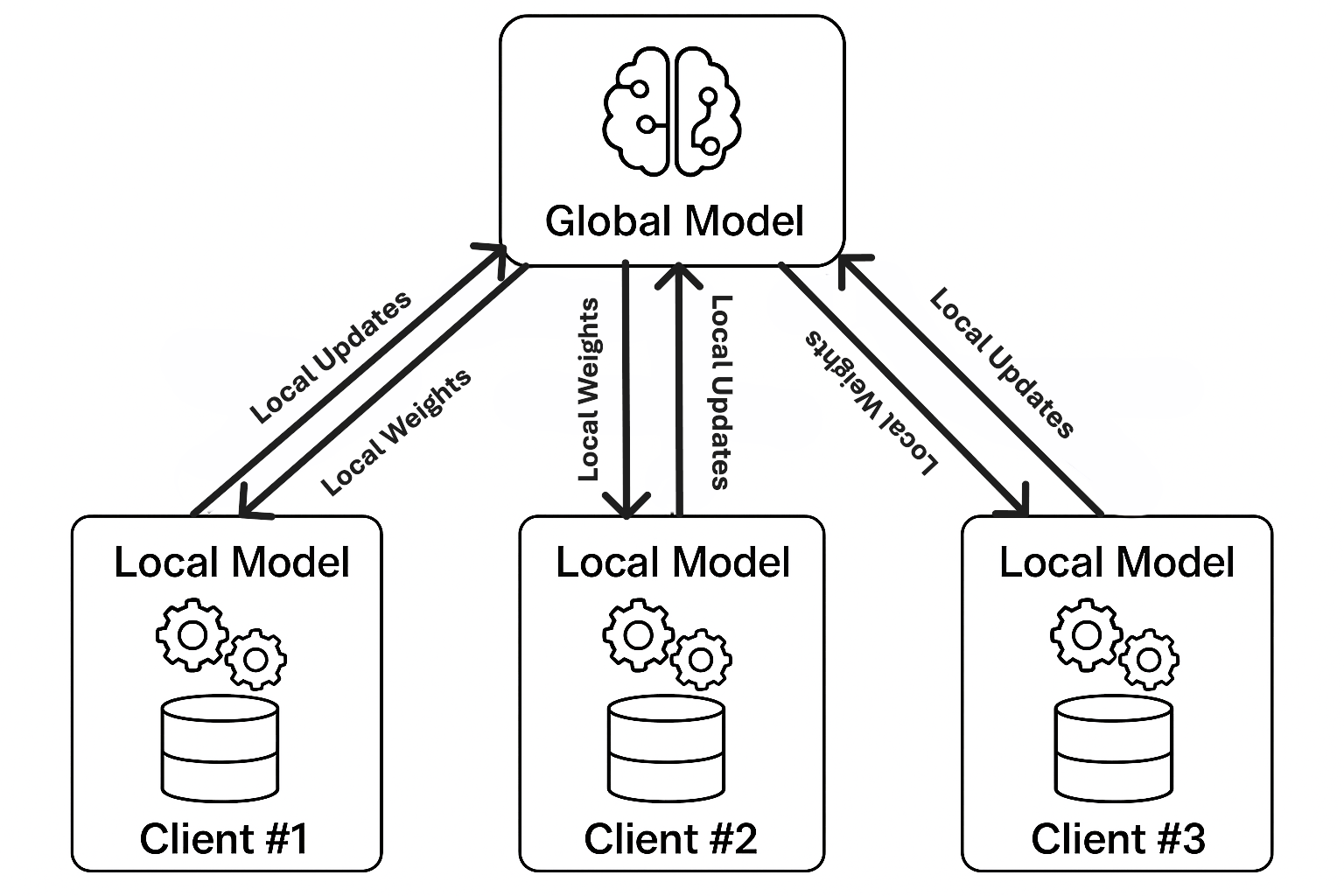}
    \caption{Federated Learning workflow for KE-NET showing parameter aggregation process.}
    \label{fig:federated_learning_kenet}
\end{figure}

Figure \ref{fig:federated_learning_kenet} illustrates the federated learning paradigm for semantic communication. Each client maintains a local dataset and trains the KE-NET model independently. Periodically, clients transmit only model parameters to a central server, which aggregates them using Federated Averaging (FedAvg) \cite{mcmahan2017fedavg} and redistributes the global model. This decentralized approach preserves data privacy, accommodates heterogeneous data distributions, and improves generalization. 
We evaluate performance using BLEU scores across varying SNR conditions to demonstrate practical benefits in realistic environments.
The central server distributes the current global parameters $\theta_{t-1}$ to all clients at each communication round. Each client $i$ performs local training using its dataset $\mathcal{D}_i$ and knowledge graph $\mathcal{G}_i$, updating the model by minimizing the local loss function:
\begin{equation}
    \theta_i^t = \arg\min_{\theta} \mathcal{L}_{\text{KE}}(\theta; \mathcal{D}_i, \mathcal{G}_i)
    \label{Eq1}
\end{equation}
After local updates, clients send their parameters $\theta_i^t$ to the server, which aggregates them using the Federated Averaging  algorithm~\cite{mcmahan2017fedavg}:
\begin{equation}
    \theta_t = \sum_{i=1}^K w_i \theta_i^t, \quad w_i = \frac{|\mathcal{D}_i|}{\sum_{j=1}^K |\mathcal{D}_j|}
\end{equation}
The global objective is to minimize the weighted sum of local losses across all clients~\cite{mcmahan2017fedavg,kairouz2021advances}:
\begin{equation}
    \min_{\theta} \sum_{i=1}^K w_i \, \mathcal{L}_{\text{KE}}(\theta; \mathcal{D}_i, \mathcal{G}_i)
    \label{EQ1}
\end{equation}
This approach enables robust model generalization while ensuring that raw data remains decentralized.

\subsection{Gradient-Based Recursions for Global Objective Minimization}

To minimize the global objective function in Equation \eqref{EQ1}, we derive the gradient-based recursions. For each client $i$, the local gradient update follows the standard gradient descent rule:
\begin{equation}
\theta_i^{t,k+1} = \theta_i^{t,k} - \eta \nabla_{\theta} \mathcal{L}_{\text{KE}}(\theta_i^{t,k}; \mathcal{D}_i, \mathcal{G}_i)
\end{equation}
where $\theta_i^{t,k}$ represents the parameters at client $i$ during communication round $t$ and local iteration $k$, and $\eta$ is the learning rate. Alternatively, many adaptive techniques \cite{aifir,jio,jidf} could be used to train the clients.

Since KE-NET uses a multi-objective approach, the total loss function consists of four components:
\begin{equation}
\mathcal{L}_{\text{KE}}(\theta; \mathcal{D}_i, \mathcal{G}_i) = \alpha \mathcal{L}_{\text{semantic}} + \beta \mathcal{L}_{\text{knowledge}} + \gamma \mathcal{L}_{\text{channel}} + \delta \mathcal{L}_{\text{reconst}}
\end{equation}
The gradient computation for the multi-objective loss is:
\vspace{-0.5em}
\begin{align}
\nabla_{\theta}\mathcal{L}_{\text{KE}}(\theta_i^{t,k})
= \alpha \nabla_{\theta}\mathcal{L}_{\text{semantic}}
+ \beta \nabla_{\theta}\mathcal{L}_{\text{knowledge}}\nonumber \\
+ \gamma \nabla_{\theta}\mathcal{L}_{\text{channel}}
+ \delta \nabla_{\theta}\mathcal{L}_{\text{reconst}}
\end{align}
\vspace{-0.5em}
After $E$ local epochs, each client's parameters become: 
\begin{equation}
\theta_i^{t+1} = \theta_i^{t,E} = \theta_t - \eta \sum_{k=0}^{E-1} \nabla_{\theta} \mathcal{L}_{\text{KE}}(\theta_i^{t,k}; \mathcal{D}_i, \mathcal{G}_i)
\end{equation}
The global parameter update through Federated Averaging FedAvg incorporates all loss components:

\vspace{-1em}
\begin{equation}
\theta_{t+1} = \sum_{i=1}^K w_i \theta_i^{t+1}
\end{equation}
Substituting the local updates, we obtain

\vspace{-1em}
\begin{equation}
\begin{aligned}
\theta_{t+1}
=\, \theta_t 
&- \eta \sum_{i=1}^K w_i \sum_{k=0}^{E-1} \Big(
\alpha \nabla_{\theta} \mathcal{L}_{\text{semantic}}(\theta_i^{t,k}; \mathcal{D}_i) \\
&\quad + \beta \nabla_{\theta} \mathcal{L}_{\text{knowledge}}(\theta_i^{t,k}; \mathcal{G}_i)
+ \gamma \nabla_{\theta} \mathcal{L}_{\text{channel}}(\theta_i^{t,k}; \mathcal{D}_i) \\
&\quad + \delta \nabla_{\theta} \mathcal{L}_{\text{reconst}}(\theta_i^{t,k}; \mathcal{D}_i)
\Big). 
\end{aligned}
\end{equation}
\vspace{-1em}
\subsection{Detailed Gradient Expressions}

Since BLEU is non-differentiable, we optimize a Minimum Risk Training (MRT) objective \cite{STE}, where gradients are computed from the differentiable expected risk rather than the BLEU score itself. The following expression provides an approximate semantic optimization objective:
\begin{equation}
\nabla_{\theta} \mathcal{L}_{\text{semantic}} = -\frac{1}{N} \sum_{n=1}^{N} w_n \left[
\frac{1}{BP} \cdot \frac{\partial BP}{\partial \theta}
+ \frac{1}{n} \sum_{j=1}^{n} \frac{1}{p_j} \cdot \frac{\partial p_j}{\partial \theta}
\right]
\end{equation}

The knowledge graph loss gradient follows:
\begin{equation}
\nabla_{\theta} \mathcal{L}_{\text{knowledge}} = \sum_{j=1}^{N_t} -w_j \left[
\frac{t_j}{\hat{t}_j} - \frac{1 - t_j}{1 - \hat{t}_j}
\right] \cdot \frac{\partial \hat{t}_j}{\partial \theta}
\end{equation}

The reconstruction loss gradient is:
\begin{equation}
\nabla_{\theta} \mathcal{L}_{\text{reconst}} = \frac{2}{|S|}\sum_{j=1}^{|S|} (\hat{\mathbf{s}}_j - \mathbf{s}_j)^T \frac{\partial \hat{\mathbf{s}}_j}{\partial \theta}
\end{equation}

The channel-aware loss gradient accounts for channel impairments:
\begin{equation}
\nabla_{\theta} \mathcal{L}_{\text{channel}} = 2\,\mathbb{E}_{\mathbf{h}\sim p(\mathbf{h})}\left[(\mathbf{h} \odot \hat{\mathbf{y}} - \mathbf{y})^T\left(\mathbf{h}\odot\frac{\partial \hat{\mathbf{y}}}{\partial \theta}\right)\right]
\end{equation}

This multi-objective gradient recursion ensures that the federated learning process optimizes semantic encoding quality, knowledge consistency, channel robustness that can be enhanced with channel coding \cite{memd,bfpeg,dopeg,vfap,kaids}, and reconstruction accuracy simultaneously while preserving data privacy across distributed clients.

\section{Theoretical Analysis: SNR Enhancement through Federated Aggregation}
\label{section4}
This section analyzes the SNR gain obtained when aggregating signals from $N$ distributed clients under independent noise conditions.
For a single client with signal power $P_{\text{signal}}$ and noise power $P_{\text{noise}}$: \vspace{-1.5em}
\begin{equation}
    \text{SNR}_{\text{client}} = \frac{P_{\text{signal}}}{P_{\text{noise}}}
\end{equation}
The server receives $N$ signals: $s_1 + n_1, s_2 + n_2, \ldots, s_N + n_N$ and averages them: \vspace{-1em}
\begin{equation}
    \text{Aggregate signal} = \frac{1}{N} \sum_{i=1}^N (s_i + n_i)
\end{equation} 
After averaging, noise power reduces while signal power remains constant:
\vspace{-1em}
\begin{equation}
P_{\text{noise, server}} = \frac{P_{\text{noise}}}{N}, \quad P_{\text{signal, server}} = P_{\text{signal}}
\end{equation}
\vspace{-1em}
This results in enhanced SNR as described by
\begin{equation}
\text{SNR}_{\text{server}} = N \cdot \text{SNR}_{\text{client}}
\end{equation}
In decibels, we have
\begin{equation}
\text{SNR}_{\text{dB, server}} = \text{SNR}_{\text{dB, client}} + 10 \log_{10}(N)
\end{equation}
\vspace{-2em}
\section{Experimental Results and Analysis}
\label{section5}
This section presents the theoretical analysis of SNR gains achieved through federated learning aggregation and contrasts these results with baseline KE-NET, GenSC, KER, DeepSC, and (5 bit + Turbo) transmission models.

\subsection{Dataset Description}
We employ 50,000 sentences from news articles \cite{ag_news}(40\%), social media  \cite{sentiment140} (35\%), and technical documents  \cite{arxiv_dataset}(25\%), with sentence lengths ranging from 5-50 tokens (average: 15.3) and a vocabulary of 25,000 tokens. The knowledge base integrates ConceptNet 5.7 \cite{conceptnet57} and WordNet 3.1 \cite{wordnet1995}, totaling 2.5 million entities and 8 million relationships.Current experiments use one dataset and fixed client partitions. Broader multi-dataset and larger-scale federated evaluations are ongoing.

\vspace{-1em}
\subsection{Federated Data Distribution}
The dataset is partitioned into 10 equal segments of 5,000 sentences each in an IID (Independent and Identically Distributed) manner, maintaining consistent domain distribution across all partitions. Each partition is split into training (70\%), validation (15\%), and test (15\%) subsets. 
\vspace{-.5em}
\begin{table}[ht]
\centering
\caption{Federated Learning Data Distribution}
\begin{tabular}{|c|c|c|c|c|}
\hline
\textbf{Config} & \textbf{Clients} & \textbf{Training} & \textbf{Validation} & \textbf{Test} \\
 & \textbf{($N$)} & \textbf{(total)} & \textbf{(total)} & \textbf{(total)} \\
\hline
1 & 1 & 3,500 & 750 & 750 \\
2 & 2 & 7,000 & 1,500 & 1,500 \\
3 & 4 & 14,000 & 3,000 & 3,000 \\
4 & 10 & 35,000 & 7,500 & 7,500 \\
\hline
\end{tabular}
\label{tab:fl_distribution}
\end{table}
\vspace{-1em}


\def\awgnsemantickenet{(0,0.579) (3,0.760) (6,0.850) (9,0.910) (12,0.940) (15,0.960) (18,0.970)}
\def\awgnsemanticgensc{(0,0.520) (3,0.680) (6,0.790) (9,0.860) (12,0.900) (15,0.920) (18,0.930)}
\def\awgnsemanticdeepsc{(0,0.320) (3,0.420) (6,0.520) (9,0.620) (12,0.680) (15,0.720) (18,0.750)}
\def\awgnsemantictradi{(0,0.00) (3,0.02) (6,0.06) (9,0.15) (12,0.28) (15,0.42) (18,0.52)}
\def\fadingsemantickenet{(0,0.4246) (3,0.5859) (6,0.6888) (9,0.7753) (12,0.8170) (15,0.8598) (18,0.8952)}
\def\fadingsemanticgensc{(0,0.3481) (3,0.4786) (6,0.5845) (9,0.6690) (12,0.7142) (15,0.7523) (18,0.7837)}
\def\fadingsemanticdeepsc{(0,0.2040) (3,0.2815) (6,0.3664) (9,0.4593) (12,0.5139) (15,0.5607) (18,0.6019)}
\def\fadingsemantictradi{(0,0.00) (3,0.01) (6,0.04) (9,0.12) (12,0.25) (15,0.36) (18,0.50)}

\def\awgnsingleclient{(0,0.8933) (3,0.9101) (6,0.9537) (9,0.9735) (12,0.9698) (15,0.9728) (18,0.9737)}
\def\awgn10clients{(0, 0.9599) (3,0.9774) (6,0.9779) (9,0.9779) (12,0.9780) (15,0.9780) (18,0.9780)}
\def\fadingsingleclient{(0,0.820) (3,0.880) (6,0.910) (9,0.940) (12,0.955) (15,0.965) (18,0.970)}
\def\fading10clients{(0,0.900) (3,0.940) (6,0.955) (9,0.965) (12,0.970) (15,0.975) (18,0.978)}

\def\awgnker{(0,0.512) (3,0.665) (6,0.782) (9,0.835) (12,0.885) (15,0.916) (18,0.920)}
\def\fadingker{(0,0.311) (3,0.419) (6,0.535) (9,0.618) (12,0.645) (15,0.682) (18,0.715)}


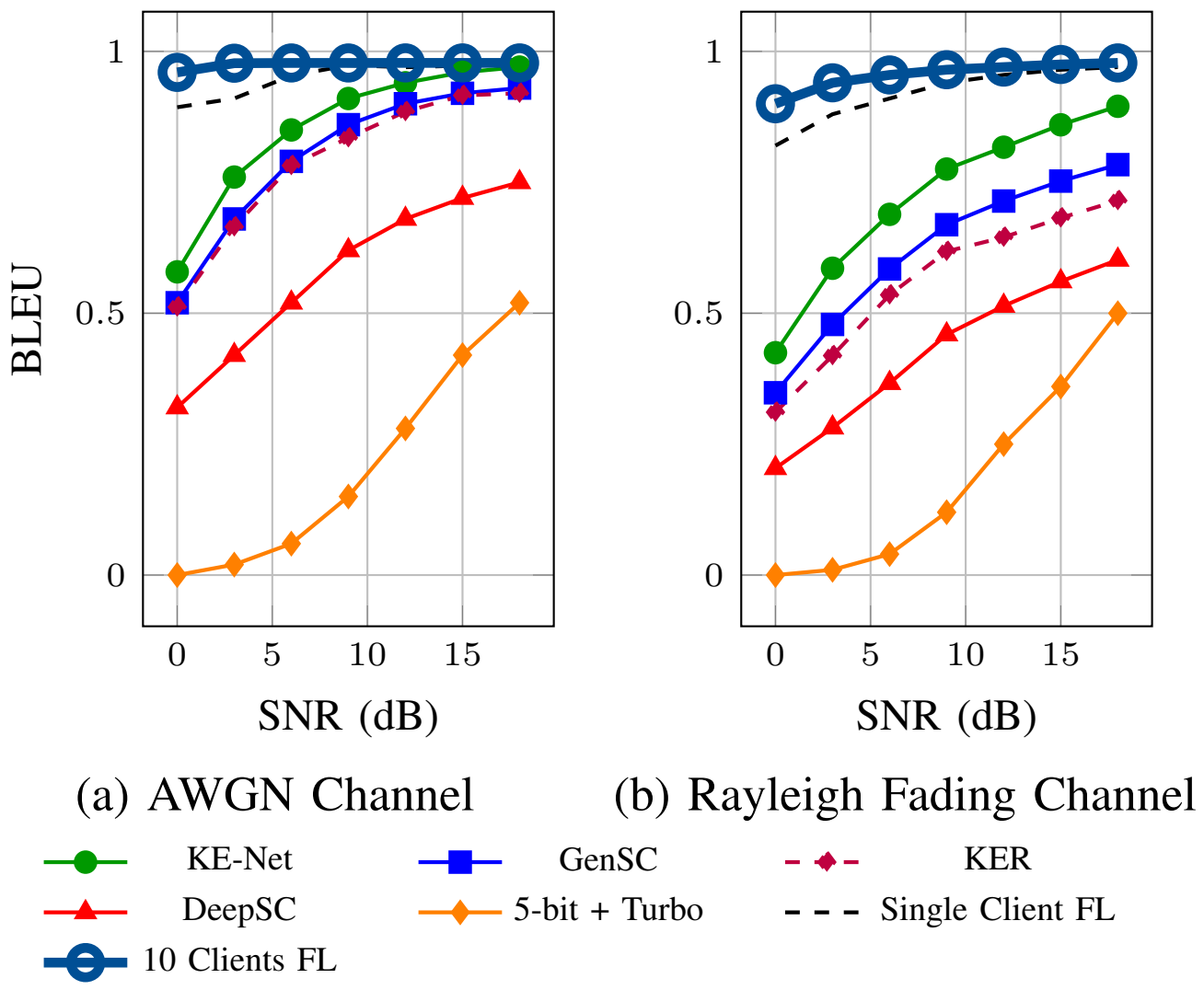
\begin{figure}[t]
\centering
\begin{tabular}{cc}

\begin{tikzpicture}
\begin{axis}[
  width=0.25\textwidth, 
  height=6cm,
  xlabel={SNR (dB)},
  ylabel={BLEU},
  tick label style={font=\scriptsize},
  label style={font=\small},
  grid=major,
  legend style={font=\scriptsize},
]

\addplot[green! 60! black, thick, mark=*] coordinates {\awgnsemantickenet};
\addplot[blue, thick, mark=square*] coordinates {\awgnsemanticgensc};
\addplot[red, thick, mark=triangle*] coordinates {\awgnsemanticdeepsc};
\addplot[orange, thick, mark=diamond*] coordinates {\awgnsemantictradi};
\addplot[black, dashed, thick, mark=star*] coordinates {\awgnsingleclient};
\addplot[mark=o, color=ieeeblue, line width=2pt, mark size=3pt] coordinates {\awgn10clients};

\addplot[purple, dashed, thick, mark=diamond*] coordinates {\awgnker};

\end{axis}
\end{tikzpicture}
&
\begin{tikzpicture}
\begin{axis}[
  width=0.25\textwidth, 
  height=6cm,
  xlabel={SNR (dB)},
  tick label style={font=\scriptsize},
  label style={font=\small},
  grid=major,
  legend style={font=\scriptsize},
]

\addplot[green!60!black, thick, mark=*] coordinates {\fadingsemantickenet};
\addplot[blue, thick, mark=square*] coordinates {\fadingsemanticgensc};
\addplot[red, thick, mark=triangle*] coordinates {\fadingsemanticdeepsc};
\addplot[orange, thick, mark=diamond*] coordinates {\fadingsemantictradi};
\addplot[black, dashed, thick, mark=star*] coordinates {\fadingsingleclient};
\addplot[mark=o, color=ieeeblue, line width=2pt, mark size=3pt] coordinates {\fading10clients};

\addplot[purple, dashed, thick, mark=diamond*] coordinates {\fadingker};

\end{axis}
\end{tikzpicture}
\\
(a) AWGN Channel & (b) Rayleigh Fading Channel
\end{tabular}

\begin{tikzpicture}
\begin{axis}[
  hide axis,
  xmin=0, xmax=1, ymin=0, ymax=1,
  legend columns=3,
  legend style={
    font=\scriptsize,
    at={(0.5,-0.1)},
    anchor=north,
    draw=none,
    /tikz/every even column/.append style={column sep=0.5cm}
  }
]

\addlegendimage{green!60!black, thick, mark=*}
\addlegendentry{KE-Net}
\addlegendimage{blue, thick, mark=square*}
\addlegendentry{GenSC}
\addlegendimage{purple, dashed, thick, mark=diamond*}
\addlegendentry{KER}
\addlegendimage{red, thick, mark=triangle*}
\addlegendentry{DeepSC}
\addlegendimage{orange, thick, mark=diamond*}
\addlegendentry{5-bit + Turbo}
\addlegendimage{black, dashed, thick, mark=star*}
\addlegendentry{Single Client FL}
\addlegendimage{mark=o, color=ieeeblue, line width=2pt, mark size=3pt}
\addlegendentry{10 Clients FL}

\end{axis}
\end{tikzpicture}

\caption{BLEU score comparison under (a) AWGN and (b) Rayleigh fading channels.}
\label{fig:bleu_comparison}
\end{figure}

\begin{figure}[ht]
\centering
\begin{tikzpicture}
\begin{axis}[
    width=8cm,
    height=6cm,
    xlabel={SNR (dB)},
    ylabel={BLEU Score},
    xmin=-20, xmax=20,
    ymin=0.25, ymax=1.05,
    grid=major,
    legend pos=south east,
    legend style={
        draw=black,
        fill=white,
        fill opacity=0.8,
        text opacity=1,
        font=\small
    },
    every axis plot/.append style={thick},
    mark size=2.  5pt,
    xtick={-20,-10,0,10,20},
    ytick={0,0.2,0.4,0.6,0.8,1.0},
]

\addplot[color=blue, mark=*, line width=1.2pt] coordinates {   
    (-20, 0.4064) (-10, 0.9527) (0, 0.9961) (10, 0.9965) (20, 0.9961) 
};
\addlegendentry{10 Clients}

\addplot[color=orange, mark=diamond*, line width=1.2pt] coordinates {
    (-20, 0.3829) (-10, 0.9386) (0, 0.9908) (10, 0.9924) (20, 0.9927)
};
\addlegendentry{4 Clients}

\addplot[color=red, mark=square*, line width=1.2pt] coordinates {
    (-20, 0.3496) (-10, 0.9193) (0, 0.9812) (10, 0.9842) (20, 0.9846)
};
\addlegendentry{2 Clients}

\addplot[color=green!  60! black, mark=triangle*, line width=1.2pt] coordinates {
    (-20, 0.3074) (-10, 0.8596) (0, 0.9589) (10, 0.9622) (20, 0.9631)
};
\addlegendentry{1 Client}

\end{axis}
\end{tikzpicture}
\caption{BLEU Score vs SNR for different numbers of federated clients showing the benefit of collaborative training.}
\label{fig:fl_clients}
\end{figure}
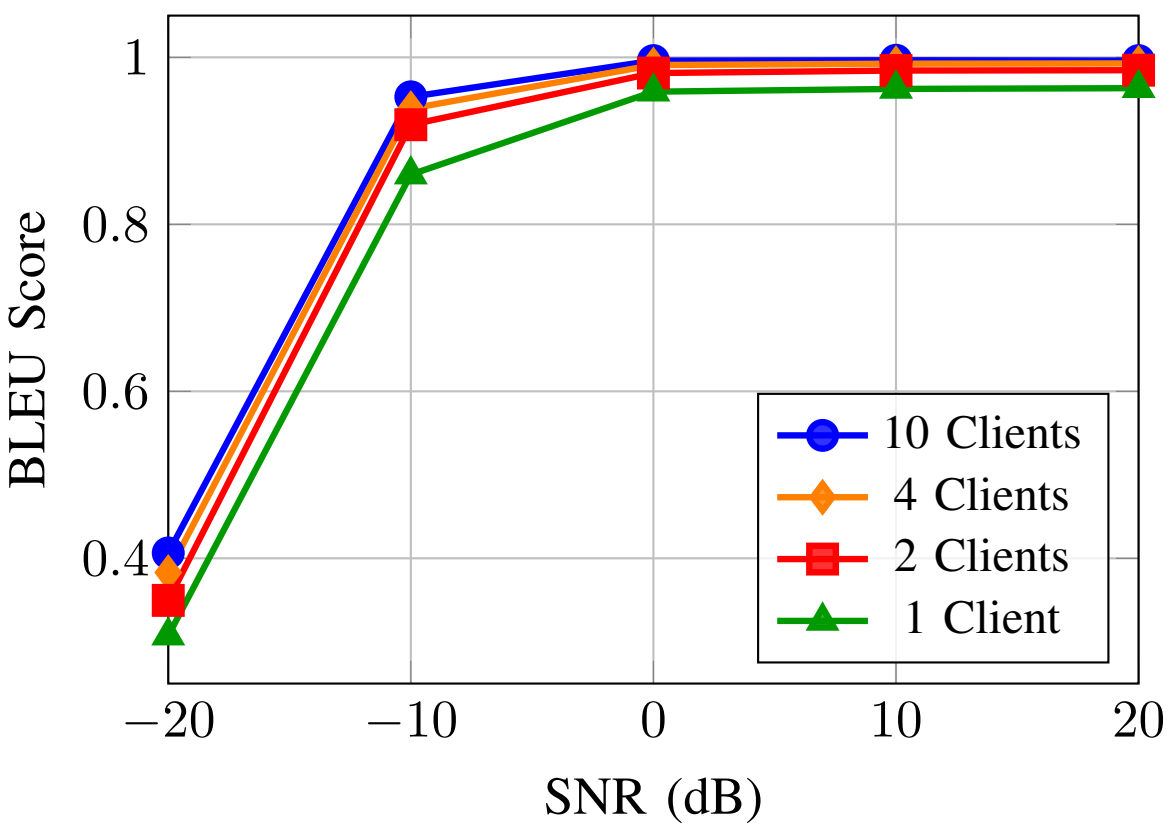 \vspace{-1em}
\subsection{Performance Analysis}

Figure~\ref{fig:bleu_comparison} compares BLEU scores across SNR for semantic communication models under AWGN and Rayleigh fading channels: KE-NET, GenSC, DeepSC, KER, conventional 5-bit + Turbo coding, single-client FL, and 10-client FL.

Results demonstrate a clear performance hierarchy: 10-client FL achieves the highest BLEU scores (~0.978), followed by single-client FL. Among centralized methods, KE-NET outperforms GenSC and DeepSC. KER performs between GenSC and KE-NET (0.92 AWGN, 0.715 fading at 18 dB). All semantic approaches substantially outperform conventional 5-bit + Turbo coding, particularly at low SNR.

The performance gap between approaches is most pronounced under challenging conditions (low SNR and fading channels), where semantic methods demonstrate their robustness. KER's performance trajectory shows particular promise in AWGN channels, where it approaches KE-NET's performance levels, suggesting that retrieval-based knowledge enhancement can be an effective alternative to direct knowledge integration. Under fading conditions, while all methods experience degraded performance, the relative advantages of federated learning and knowledge-enhanced approaches become even more apparent, with KER maintaining reasonable performance despite the challenging channel conditions.

KER demonstrates competitive performance under both channel conditions. In AWGN channels, it achieves BLEU scores ranging from 0.512 at 0 dB to 0.920 at 18 dB, approaching the performance of KE-NET. Under Rayleigh fading, performance decreases as expected due to channel impairments, increasing from 0.311 to 0.715 across the evaluated SNR range. Overall, KER consistently outperforms GenSC while remaining below KE-NET, indicating that retrieval-based knowledge enhancement provides an effective compromise between conventional semantic communication models and fully integrated knowledge-enhanced architectures.
Figure \ref{fig:fl_clients} demonstrates the impact of SNR on BLEU score performance for federated semantic communication systems with varying client populations. Results show a critical transition between -20 dB and -10 dB SNR where BLEU scores surge from 0.31-0.41 to 0.86-0.95, representing over 180 \% improvement. The 10-client configuration consistently outperforms smaller deployments, achieving 32 \% higher BLEU at -20 dB SNR compared to the single-client baseline, though this advantage diminishes to 3.5 \% at 20 dB SNR. All configurations converge to near-perfect performance ($BLEU > 0.95$) beyond 0 dB SNR, indicating that federated diversity provides maximal benefit under challenging channel conditions while architectural constraints dominate under favorable conditions.
\begin{algorithm}[ht]
\caption{Federated KE-NET Training Algorithm}
\label{alg:fl_training}
\begin{algorithmic}[1]
\Require Initial global model parameters $\theta_0$, client datasets $\{\mathcal{D}_i\}_{i=1}^K$,  communication rounds $T=15$, local epochs $E= 8$ learning rate $0.0001$
\Ensure Trained global model parameters $\theta^*$
\State Initialize global model parameters $\theta_0$
\For{round $t = 1$ to $T$}
    \For{each client $i = 1$ to $K$ \textbf{in parallel}}
        \State Receive global model $\theta_{t-1}$ from server
        \State Perform local KE-NET training:
        \State \quad $\mathcal{L}_{\text{total}} \gets \alpha \mathcal{L}_{\text{semantic}} + \beta \mathcal{L}_{\text{knowledge}} + \gamma \mathcal{L}_{\text{reconst}} + \delta \mathcal{L}_{\text{channel}}$
        \State \quad $\theta_i^t \gets \theta_{t-1} - \eta \nabla_\theta \mathcal{L}_{\text{total}}$
        \State Send updated parameters $\theta_i^t$ to server
    \EndFor
    \State Server aggregates: $\theta_t = \sum_{i=1}^K w_i \theta_i^t$
    \State Distribute updated global model $\theta_t$ to all clients
    \If{convergence criterion met}
        \State \textbf{break}
    \EndIf
\EndFor
\State \Return $\theta^*$
\end{algorithmic}
\end{algorithm}
\section{Conclusion}
\label{section6}

This paper integrates federated learning with knowledge-enhanced networks for privacy-preserving semantic communication. Federated aggregation achieves $10\log_{10}(N)$ dB SNR improvement, with strong low-SNR performance validated experimentally. Future work will investigate communication-efficient federated optimization, parameter transmission overhead, and multi-modal semantic communication.

\end{document}